\documentclass[twoside,fleqn]{article}
\usepackage{latexsym,espcrc2,epsf}
\usepackage{amsfonts}
\usepackage{epsfig}

\title{
Operator product expansion and non-perturbative
renormalization }

\author{
  Sergio Caracciolo${}^{\rm a}$, 
  Andrea Montanari \address{Scuola Normale Superiore and INFN,
      Sezione di Pisa,
      I-56100 Pisa, ITALIA},
 $\!$and
  Andrea Pelissetto\address{Dipartimento di Fisica and INFN,
      Universit\`a degli Studi di Pisa,
      I-56100 Pisa, ITALIA}
}

\begin{document}

\begin{abstract}
It has been recently proposed to use the operator product expansion to
evaluate the expectation values of renormalized operators without the
need of a direct
computation of the relevant renormalization constants. 
We test the viability of this idea in the two-dimensional
non-linear $\sigma$-model discussing the 
non-perturbative renormalization of the 
energy-momentum tensor.
\end{abstract}

\maketitle

\newcommand{\reff}[1]{(\ref{#1})}
\def\smfrac#1#2{{\textstyle\frac{#1}{#2}}}

\newcommand{\be}{\begin{equation}}
\newcommand{\ee}{\end{equation}}
\newcommand{\<}{\langle}
\renewcommand{\>}{\rangle}

\newcommand\FF{{\cal F}_\delta}

\def\spose#1{\hbox to 0pt{#1\hss}}
\def\ltapprox{\mathrel{\spose{\lower 3pt\hbox{$\mathchar"218$}}
 \raise 2.0pt\hbox{$\mathchar"13C$}}}
\def\gtapprox{\mathrel{\spose{\lower 3pt\hbox{$\mathchar"218$}}
 \raise 2.0pt\hbox{$\mathchar"13E$}}}

\def\bsigma{\mbox{\protect\boldmath $\sigma$}}
\def\bpi{\mbox{\protect\boldmath $\pi$}}
\def\btau{\mbox{\protect\boldmath $\tau$}}
\def\hatp{\hat p}
\def\hatl{\hat l}

\def\msbar{ {\overline{\hbox{\scriptsize MS}}} }
\def\normalmsbar{ {\overline{\hbox{\normalsize MS}}} }

\newcommand{\R}{\hbox{{\rm I}\kern-.2em\hbox{\rm R}}}

\def\smfrac#1#2{{\textstyle\frac{#1}{#2}}}

The Operator Product Expansion (OPE) consists in rewriting,
for $y\to x$, 
\be
A(x) B(y) \to \sum_C W_C^{AB}(x-y) C(y)\; .
\label{f1}
\ee
Here $A$, $B$, and $C$ are operators; 
$W_C^{AB}(x-y)$ are $c$-number functions (Wilson coefficients), 
singular when $|x-y|\to 0$, that
can be computed, for example, in perturbation theory.
Dimensional analysis tells us that the leading contribution
for $y\to x$ is due to 
the operators $C$'s in Eq. \reff{f1}
that have the lowest dimension.
Notice that the  
OPE is an operator relation and therefore, for any matrix
element $\<\psi| A(x)B(y) |\phi\>$, the coefficients
$W_C^{AB}(x-y)$ are independent of the states $|\psi\>$ and  $|\phi\>$.
This also means that if one wants to use the OPE,
one should go through all the subtleties in the definition of the
operators in quantum field theory (regularization,
renormalization, \dots).

The authors of reference \cite{Rossietal} have proposed to
consider, for lattice applications, the OPE in the
particular case in which $A$ and $B$ are the components of a
conserved current $J_\mu$. This case is particularly simple 
because the components of $J_\mu$
do not need to be renormalized. This means that
\be
\left[ J_\mu(x) J_\nu(0) \right]_{MS}(\mu) =
\left[ J_\mu(x) J_\nu(0) \right]_{L}({1/ a})\; ,
\label{f3}
\ee
where the currents are expressed in terms of the
bare fields and the coupling constant on the lattice
is related to the coupling constant $g$ in the continuum at
the scale $\mu$ by a renormalization factor
\be
g_{L}({1/ a}) = Z_g^{L}\left( g(\mu), \mu a \right)
\, g(\mu)\; .
\label{f4}
\ee
Therefore, the lattice computation of $\< \psi| J_\mu(x) J_\nu(0)| \phi\>$
provides the same expectation value as in the continuum, say in the
MS-scheme, at the
scale $\mu$. One can then use Eq. \reff{f1} 
to estimate the matrix elements of the
operators $C$'s that dominate the r.h.s. of \reff{f1} (in
the continuum at the scale $\mu$), 
bypassing all the troubles of lattice
perturbation theory (large deviations from 1 of the $Z$'s, large corrections,
improvements, \dots). 
The recipe is the following. One evaluates on the lattice the matrix
elements
$\<\psi| J_\mu(x)J_\nu(0) |\phi\>$ in the range $1 < |x| < \xi$ (where
$\xi$ is as usual the correlation length) and then uses the
OPE in the continuum (at the scale $\mu$) to fit the
(now) numbers $\<\psi|C(0)|\phi\>$. One needs the Wilson 
coefficients $W_C^{\mu\nu}(x)$. The idea is to use their
expression computed in
perturbation theory in the continuum.

Assuming $C(0)$ to be multiplicatively renormalized, one can then use
\be
\<\psi|C(0)|\phi\> = Z_C\left(g_{L}({1/ a}),\mu a\right)
\<\psi|C^{L}(0)|\phi\>\; .
\ee
In this way, by measuring  
$\<\psi|C^{L}(0)|\phi\>$ on the lattice, 
we obtain a non-perturbative determination of $Z_C$. 

The discussion in \cite{Rossietal} ends with a question: is it
feasible? That is, does it exist, in practice, a numerically
accessible window for $|x|$ where, for a truncation of the sum
over $C$, the OPE applies accurately enough to extract the
expectation values  of the $C$'s?

To perform a check we need a case in which we already know the 
matrix elements
$\<\psi|C(0)|\phi\>$, so that we can verify the correctness of 
our results.  

Our example, given the experience that we have accumulated
along the past years, will refer to the $O(N)$
$\sigma$-model  in two dimensions  with lattice action 
\be
S =\, {1\over 2 g_{L}} \sum_{x,\mu} \left(\partial_\mu
\bsigma \right)^2_x
\ee
with $\bsigma\in S^{N-1}$  and $\left(\partial_\mu
f\right)_x = f_{x+\mu} - f_x$.
The fields are related to their continuum version by
\be
\bsigma_x = \sqrt{Z^{L}} \bsigma(x)\;.
\ee
All the renormalization factors $Z^{L}_i$ have an
expansion of the form
\be
Z_i^{L}(g,\mu,a) = \sum_{l=0}^\infty g^l \sum _{n=0}^l
c_n^{(l)} \ln^n \mu a\; .
\ee
For example $Z_g^L$ and $Z^L$
are known up to four loops~\cite{noi}.

For our purposes we made use of the non-perturbative
determination of $Z_g^{L}$  given in~\cite{LWW} for the
case $N=3$.

\begin{figure}
\begin{center}
\epsfig{figure=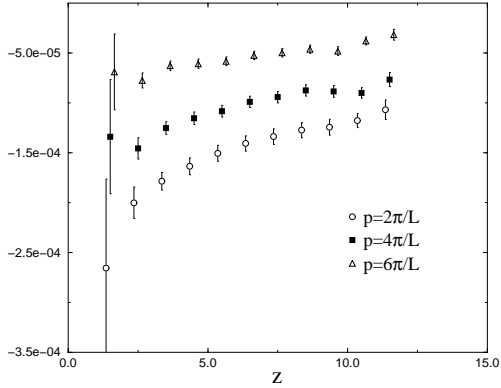,angle=-90,width=\linewidth}
\end{center}
\caption{
Angular average $g^2_L \overline{\cal I}_{01}(z;p,20)/2$. The abscissa
of coincident points have been splitted. Notice that the rightmost points
are not a faithful angular average because ${\cal I}$ has been 
measured only on a square.
}
\label{fig1}
\end{figure}

The Noether currents related to the $O(N)$-invariance, which
is preserved also on the lattice, are
\def\bj{\mbox{\protect\boldmath $j$}}
\be
j_{\mu}^{a,b}(x) = {1\over g} \left(\bsigma_x^a \partial_\mu
  \bsigma_x^b - \bsigma_x^b \partial_\mu
  \bsigma_x^a\right),
\ee
and their singlet product is
\be
\bj_{\mu}(x) \cdot \bj_{\nu}(0) = \sum_{a,b} j_{\mu}^{a,b}(x)
                j_{\nu}^{a,b}(0)\; .
\ee
In the continuum, using the OPE and 
averaging over a circle, we obtain
\begin{eqnarray}
\lefteqn{{g^2\over 2} \overline{\bj_{1}(x) \cdot \bj_{0}(0)}
=} \nonumber \\ 
&& ={g^2}\, \int {d\theta\over 4 \pi} \bj_{1}(r
\cos \theta, r\sin \theta) \cdot \bj_{0}(0,0) \nonumber\\
&& =\left[\partial_1 \bsigma \cdot \partial_0
\bsigma\right]_{MS}(0)\, W_C^{10}(r) +O(r^2)\nonumber\\
&& =g\,T_{10}^{MS}(0) \, W_C^{10}(r) +O(r^2)
\label{op}
\end{eqnarray}
where $T^{MS}$ is the continuum energy-momentum tensor (EMT), which is
exactly conserved, a  property that is not shared by its lattice
counterpart. It follows that
\be
\int T_{10}^{MS}(x,t) \,dx = p,
\ee
where $p$ is the momentum, so that, in a strip of size $L$, the expectation
value between states of momentum $p$ is 
\be
{\<p|T_{10}^{MS}(0)|p\> \over \<p|p\>} = {p\over L}.
\ee
The Wilson coefficient is given  by
\begin{eqnarray}
W_C^{10}(r) &=& 1 + {5(N-2)\over 8 \pi}\, g   \\
&& - {N-2\over 4 \pi}\,g \,(\gamma + \ln \pi^2 \mu^2 r^2)+O(g^2).
\nonumber 
\end{eqnarray}
We have performed 
a Monte Carlo simulation on a lattice of
size $L\times T= 128\times 256$ with periodic boundary
conditions at $g_L^{-1} = 1.54$, which
corresponds to $\xi = 13.632(6)$. We measure:

\noindent a) the correlation function in the one-particle sector
with momentum $p$
\begin{eqnarray}
C(p,2t) &=& {1\over L} \sum_{x,y} \< \bsigma_{-t,x} \cdot
\bsigma_{t,y} \> e^{ip(x-y)} 
\nonumber \\
&\approx&  {Z(p)\over 2 \omega(p) } e^{-2t \omega(p)}; 
\end{eqnarray}
b) the correlation function of the out-of-diagonal entry
of the lattice EMT with two one-particle operators
\begin{eqnarray} 
\lefteqn{T_{10} (p,2t)\, =} \nonumber \\
 &=& {1\over g_L}{1\over L} \sum_{x,y}
\<(\overline{\partial}_1 \bsigma_{0,0}\cdot  
   \overline{\partial}_0 \bsigma_{0,0}) 
  (\bsigma_{-t,x} \cdot \bsigma_{t,y} )\> 
\nonumber \\
 & & \times e^{ip(x-y)}
\nonumber\\ &\approx&  {Z(p) \over 2 \omega(p)  }
{\<p|T_{10}^L(0)|p\> \over\<p|p\>}e^{-2t \omega(p)}; 
\end{eqnarray}
here we used the symmetric definition of the derivative
\be
\left( \overline{\partial}_\mu f \right)_x = {f_{x+\mu} -
f_{x-\mu} \over 2};
\ee
c) the correlation between the products of two currents in
the singlet sector and two one-particle operators
\begin{eqnarray}
\lefteqn{{\cal I}_{\mu\nu}  (z;p,2t)\, =} 
\nonumber \\
 &=& {1\over L} \sum_{x,y}
\<(\bj_{\mu,0} \cdot  \bj_{\nu,z}) 
  (\bsigma_{-t,x} \cdot \bsigma_{t,y} )\> e^{ip(x-y)}
\nonumber\\ &\approx&  {Z(p) \over 2 \omega(p)  } {\<p|\bj_{\mu,0}
\cdot  \bj_{\nu,z} |p\> \over
\<p|p\>}e^{-2t \omega(p)}. 
\end{eqnarray}
Moreover we need a lattice version of the angular average
\begin{eqnarray}
\overline{f}(r) &=& {1\over N(r)} \sum_z f(z)\, \Xi (z) \nonumber \\
N(r) &=& \sum_z  \Xi  (z) \nonumber \\
\Xi (z) &=& \theta\left(|z|-r+{1\over 2}\right)
\theta\left(r+{1\over 2}-|z|\right) 
\end{eqnarray}
Let us now discuss the Monte Carlo data. 
In Fig.~\ref{fig1} we plot the angular average 
$g^2_L \overline{\cal I}_{01}(z;p,20)/2$
for various distances $z$ and various momenta $p= n\, 2\pi/L$ for $n=1,2,3$.

\begin{figure}[t]
\begin{center}
\epsfig{figure=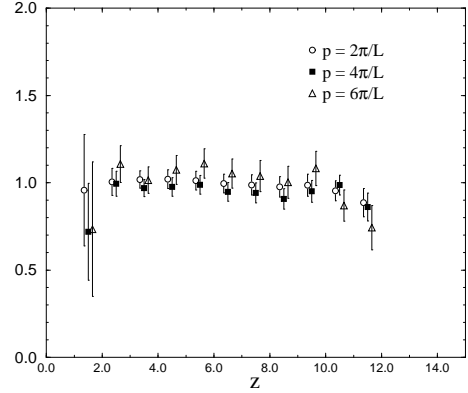,angle=-90,width=\linewidth}
\end{center}
\caption{Angular average $g^2_L \overline{\cal I}_{01}(z;p,20)/2$
divided by the theoretical prediction: 
$g\cdot$  (Wilson coefficient) $\cdot$(energy-momentum tensor). 
Here in the Wilson function the first two leading logs have been
summed up.
}
\label{fig2}
\end{figure}

In Fig.~\ref{fig2} we plot the left-hand side of \reff{op} divided by
$g T_{10}^{MS}(0) W_C^{10}(r)$
renormalized at the scale $\overline{\mu}^{-1} = 2.8519 a$. 
Here in the Wilson function the first two leading logs have been summed up.
We find 
a large window of distances for which the OPE is verified,
giving the correct matrix element independently of 
the external states. This window seems to extend even at a
distance of order $\xi/2$ which is presumably only a
lucky accident.

We also looked at the OPE directly on the lattice, i.e.
using the Wilson coefficient computed in lattice perturbation theory
resumming the first two leading logs,
the expectation value $T_{10}(p, 2t)$, and the non-perturbative
renormalization constant of the lattice EMT.
In this case we do not observe a reasonable window where the OPE
is verified.
  
We wish to thank G.~Martinelli and G.C.~~Rossi for useful discussions
on this work.

%
%
%


\begin{thebibliography}{9}

\bibitem{Rossietal}
 C.~Dawson, G.~Martinelli, G.~C.~Rossi,
C.~T.~Sachrajda, S.~Sharpe, M.~Talevi and M.~Testa,
Nucl. Phys. B514 (1998) 313.

\bibitem{noi}
S.~Caracciolo and A.~Pelissetto,
Nucl. Phys. B455 (1995) 619.


\bibitem{LWW}
 M. L\"uscher, P. Weisz and U. Wolff, 
Nucl. Phys. B359 (1991) 221.


\end{thebibliography}
\end{document}